\documentclass[%
reprint,
amsmath,amssymb,
aps,
prd,
calrsfs,
booktabs,
subcaption,
url
]{revtex4-2}

\usepackage[caption=false]{subfig}
\usepackage{amsmath, calrsfs, booktabs, url, hyperref, graphicx,ORCIDinREVTeX}
\DeclareMathAlphabet{\pazocal}{OMS}{zplm}{m}{n}

\renewcommand{\selectlanguage}[1]{}

\begin{document}

\title{Optimal Strategies for Gravitational Wave Memory Detection in Pulsar Timing Arrays}

\author{Jerry P. Sun}
\orcid{0000-0002-7933-493X}
\affiliation{Department of Physics, Oregon State University, Corvallis, OR 97331, USA}
\author{Xavier Siemens}
\orcid{0000-0002-7778-2990}
\affiliation{Department of Physics, Oregon State University, Corvallis, OR 97331, USA}
\affiliation{Center for Gravitation, Cosmology and Astrophysics, Department of Physics, University of Wisconsin-Milwaukee,\\ P.O. Box 413, Milwaukee, WI 53201, USA}
\author{Dustin R. Madison}
\orcid{0000-0003-2285-0404}
\affiliation{Department of Physics, University of the Pacific, 3601 Pacific Avenue, Stockton, CA 95211, USA}


\begin{abstract}
In this work we derive two computationally efficient frequentist detection statistics that can be used in searches for gravitational-wave bursts with memory in pulsar timing data. By maximizing the likelihood ratio in two different ways we construct a coherent statistic and an incoherent statistic, which are analogs of the $\pazocal{F}_e$ and $\pazocal{F}_p$ statistics commonly used for continuous-wave searches in pulsar timing data. We show that both statistics are $\chi^2$-distributed with varying degrees of freedom and non-centrality parameters given by the signal-to-noise (SNR) ratio of the signal present in our data. The statistics can also be used to compute the maximum likelihood estimators of amplitude parameters of a possible gravitational-wave memory signal in pulsar timing data. We find that in the low-signal regime ($\mathrm{SNR} \lesssim 5$), the estimators are inaccurate. However, in intermediate- to high-signal regimes, we show that these estimators can accurately determine the correct signal parameters. 
\end{abstract}

\maketitle

\section{Introduction}

Pulsar timing arrays (PTAs) have been very successful at probing the low-frequency part of the gravitational wave (GW) spectrum. There are many ongoing efforts to search for a diverse set of GW signals including continuous GWs from super massive black hole binary (SMBHB) systems \citep{zhu_all-sky_2014, falxa_searching_2023, agazie_nanograv_2023-2, antoniadis_second_2024}, a stochastic background of GWs from ensembles of SMBHBs \citep{antoniadis_international_2022, reardon_search_2023, agazie_nanograv_2023-1, epta_collaboration_and_inpta_collaboration_second_2023,xu_searching_2023}, nonlinear GW memory \citep{wang_searching_2015, agazie_nanograv_2023}, non-Einsteinian GW polarization modes \citep{arzoumanian_nanograv_2021}, and short-duration GW bursts from various sources \citep{becsy_bayesian_2021, deng_searching_2023}. Pulsar timing experiments use the fact that millisecond pulsars (MSPs) are extremely stable rotators \citep{lorimer_binary_2008}. The regularity of the radio pulses emitted by these pulsars make it possible to use changes in the pulse times of arrival (TOAs) to detect or set limits on various GW phenomena and perform tests of Einsteinian relativity \citep{sazhin_opportunities_1978, detweiler_pulsar_1979,hellings_upper_1983}

One such test is the search for GW bursts with memory (BWMs). As a SMBHB system completes the final phase of a merger, the burst of gravitational waves radiated through the event act as a source of so-called ``nonlinear" GW memory \citep{zeldovich_radiation_1974, christodoulou_nonlinear_1991, wiseman_christodoulous_1991, blanchet_hereditary_1992, thorne_gravitational-wave_1992}. This phenomenon is called ``nonlinear" because it originates from the nonlinearity of Einstein's field equations. When this memory wave front crosses the line of sight between the Earth and an MSP, it has the effect of changing the observed rotational frequency of this MSP. Currently, there are efforts by PTAs and ground-based GW observatories to detect nonlinear memory \cite{lasky_detecting_2016, agazie_nanograv_2023}. The work in Refs.~\cite{van_haasteren_gravitational-wave_2010, cordes_detecting_2012, madison_assessing_2014, islo_prospects_2019} also shows the detection prospects for nonlinear memory in space-based, and PTA GW experiments.

Several methods have been developed for searches for nonlinear GW memory across all GW-frequency regimes. PTA experiments largely use Markov Chain Monte Carlo sampling methods to compute posterior probabilities of memory model parameters. Because these methods are computationally intensive, we previously~\cite{sun_implementation_2023} proposed an efficient method using pre-computed likelihood tables to numerically compute posteriors on nonlinear memory model parameters.

In this work, we present two frequentist statistics which can be used in searches for nonlinear GW memory by PTAs. These statistics are optimal in the sense that they are derived by maximizing the ratio of the likelihoods of a nonlinear GW signal event to the null signal model, following the optimal frequentist strategies laid out for the stochastic GW background~\cite{anholm_optimal_2009}, and for continuous GWs~\cite{ellis_optimal_2012}. These statistics can also be used to compute estimators for the parameters that determine the amplitude of the signal. This work builds on~\cite{ellis_optimal_2012}, in which analogous statistics are derived for searches for continuous GWs. In turn, the results in \cite{ellis_optimal_2012} built on previous work~\cite{jaranowski_data_1998,cornish_search_2007,babak_resolving_2012} on frequentist statistics analytically maximized over waveform parameters for continuous GWs in LIGO, LISA, and PTA data. Finally, \cite{arzoumanian_nanograv_2015,wang_searching_2015} derived the maximum likelihood estimator for the strain amplitude of a GW memory event in PTA data. In this work we re-parameterize the signal differently allowing us to compute maximum likelihood estimators for the polarization, the amplitude, and, for the incoherent statistic, the sky location of the memory event.  

In previous work for continuous GWs~\cite{jaranowski_data_1998, cornish_search_2007,ellis_optimal_2012}, the statistic derived by maximizing the likelihood ratio was referred to as the $\pazocal{F}$-statistic. Ref. \cite{ellis_optimal_2012} further split this into a coherent statistic which they call the Earth-term $\pazocal{F}$-statistic ($\pazocal{F}_e$), and an incoherent statistic called the pulsar-term $\pazocal{F}$-statistic ($\pazocal{F}_p$) depending on how the likelihood ratio was maximized. In this work, for GW memory events, we will refer to the two different statistics that result from the two maximizations as the coherent $\pazocal{F}$-statistic ($\pazocal{F}_C$), and the incoherent $\pazocal{F}$-statistic ($\pazocal{F}_I$).  Despite progress in~\cite{sun_implementation_2023} to increase the efficiency of full Bayesian searches for nonlinear GW memory, such searches remain very computationally expensive. These new $\pazocal{F}$-statistics are useful additions to the standard set of tools used in pulsar timing data analysis. They are very fast to compute and can be used to independently cross-validate a full Bayesian search without significant additional computational costs. We expect a noise-marginalized approach analogous to that used for GW stochastic background searches \citet{vigeland_noise-marginalized_2018} to be necessary for robust GW memory searches using our frequentist techniques. We will develop these techniques in future work.

\section{Background}
In this section, we begin by reviewing the signal model for nonlinear GW memory in pulsar-timing data. We then apply a similar framework to that presented in \citet{ellis_optimal_2012} for CWs to nonlinear GW memory and develop both coherent and incoherent frequentist statistics by maximizing the likelihood ratio in two different ways.

\subsection{Signal Model}

 As a GW memory front passes through the Earth or a pulsar, the apparent rotational frequency of the pulsar changes by a fractional amount proportional to the strain-amplitude of the memory. This sudden mismatch between the pulsar's modeled rotational frequency and apparent rotational frequency induces a linear drift in the residuals of the pulsar's TOAs. This occurs because the change in rotational frequency is a constant, and we accrue a constant timing residual with every subsequent rotation of the pulsar. In this paper we consider the effect of a GW memory front passing over only the Earth, the so-called Earth-term. The rationale for this assumption is explained in detail below. The residuals induced by the Earth-term GW memory signal in the $a$-th pulsar of a PTA may be written as
\begin{widetext}
\begin{align}
    r_a(t; t_0, \mathbf{\hat{\Omega}}, h_0, \psi) = h_0 \Theta(t - t_{0}) (t - t_0)  (F^{a}_{+}(\mathbf{\hat{\Omega}})\cos{(2 \psi)} + F^{a}_{\times}(\mathbf{\hat{\Omega}})\sin{(2\psi)}),
\label{eqn:bwm_residuals}
\end{align}
\end{widetext}
where $h_0$ is the intrinsic strain of the memory signal, $t_0$ is the time at which the memory wavefront passes over the Earth, $\hat{\Omega}$ is the location of the source of the GW memory, and $\psi$ is the polarization angle, the angle between the principal polarization vector and pulsar line of sight projected onto the plane perpendicular to the propagation direction of the wave. The antenna pattern functions $F_{+}(\hat{\Omega})$ and $F_\times(\hat{\Omega})$ for plus- and cross-polarized GWs are given by
\begin{equation}
    F_{a,+}(\mathbf{\hat{\Omega}}) = \frac{1}{2} \frac{(\mathbf{\hat{m}}\cdot\mathbf{\hat{p}}_{a})^2 - (\mathbf{\hat{n}}\cdot\mathbf{\hat{p}}_{a})^2}{1 + \mathbf{\hat{\Omega}}\cdot\mathbf{\hat{p}}_{a}},
\end{equation}
\begin{equation}
    F_{a,\times}(\mathbf{\hat{\Omega}}) = \frac{1}{2} \frac{(\mathbf{\hat{m}}\cdot\mathbf{\hat{p}}_{a})(\mathbf{\hat{n}}\cdot\mathbf{\hat{p}}_{a})}{1 + \mathbf{\hat{\Omega}}\cdot\mathbf{\hat{p}}_{a}},
\end{equation}
where $\mathbf{\hat{p}}_a$ is the unit vector that points to the $a$-th pulsar. $\mathbf{\hat{m}}$ and $\mathbf{\hat{n}}$ are two orthogonal vectors that define the plane perpendicular to the propagation direction of the memory wavefront and are given by
\begin{align}
    \mathbf{\hat{m}}& = \sin\phi \mathbf{\hat{x}} - \cos\phi \mathbf{\hat{y}}\\
    \mathbf{\hat{n}}& = -(\cos\theta \cos\phi) \mathbf{\hat{x}} - (\cos\theta\sin\phi) \mathbf{\hat{y}} + (\sin\theta) \mathbf{\hat{z}},
\end{align}
where $\theta$ and $\phi$ are the polar and azimuthal angles of the source.


It is worth pointing out that the choice of only including the Earth-term GW memory signal is well motivated. The distances from the Earth to pulsars in a PTA, and their distances from one another, are on the order of hundreds to thousands of light-years. Therefore, the time it takes a GW memory front to pass over both one pulsar and the Earth, or over two different pulsars in a PTA, is hundreds to thousands of years. This timescale is much longer than typical pulsar timing experiment durations which are on the order of a decade. This means we do not expect to observe the same GW memory event passing the Earth and a pulsar, or two different pulsars in a PTA, during the course of typical pulsar timing experiments. 

It is difficult to make a compelling case for detection using single-pulsar GW memory measurements ~\citep{van_haasteren_gravitational-wave_2010, cordes_detecting_2012}. This is because we observe sudden changes, called ``glitches"~\citep{haskell_models_2015}, in the rotational frequencies of some pulsars which produce a signal in the timing residuals identical to a GW memory burst. These glitches have also been observed in MSPs, albeit very rarely~\citep{mckee_glitch_2016}.

Therefore only when a GW memory front passes over the Earth, affecting the residuals of all pulsars in the PTA in a correlated way, are we in a position to make a compelling case for detection. Individual pulsar GW memory searches, however, can still be used to place upper limits.


\subsection{Likelihood}

The total signal and noise model for a pulsar's residuals $\boldsymbol\delta \mathbf{t}$ may be written as
\begin{align}
    \boldsymbol\delta \mathbf{t} = \boldsymbol{\delta}\mathbf{t_{\mathrm{bwm}}} + \mathbf{n},
\end{align}
where $\boldsymbol{\delta}\mathbf{t_{\mathrm{bwm}}}$ are the residuals induced by the memory signal, and $\mathbf{n}$ is a timeseries containing red and white Gaussian noise. It is important to include red noise in $\mathbf{n}$
for our analyses since recent PTA data sets have been found to contain a strong GW background red noise process in addition to pulsar-intrinsic red and white noises \citep{arzoumanian_nanograv_2020, agazie_nanograv_2023-1}. The red noise paramaterized using the amplitude and spectral index $A_{\mathrm{RN}}$ and $\gamma$ via \citep{phinney_practical_2001}
\begin{align}
P_{\mathrm{RN}}(f) = A_{\mathrm{RN}}^2 \left(\frac{f}{f_{\mathrm{1yr}}}\right) ^ {-\gamma},
\label{eqn:powerlaw}
\end{align}
where $A_{\mathrm{RN}}$ is the dimensionless noise amplitude at a reference frequency of $f_{\mathrm{1yr}} = 1\mathrm{yr}^{-1}$ and $\gamma$ is the spectral index. For a sense of the expected magnitude of these parameters, current results from \citet{agazie_nanograv_2023-1} give estimates for the amplitude and spectral index of $A_{\mathrm{RN}} = 6.4^{+4.2}_{-2.7} \times 10^{-15}$ and $\gamma = 3.2^{+0.6}_{-0.6}$ for the gravitational wave background.
Ref.~\citep{arzoumanian_nanograv_2016} contains a detailed explanation of the parameterization of red and white noises in PTA data analysis. 

The residuals are obtained from the TOAs by subtracting out a timing model. To account for the effects of this subtraction we use the R-matrix formalism described in refs. \cite{demorest_limits_2013,Ellis:2013nrb}, including only the quadratic terms which originate from fitting for the rotational phase, frequency, and frequency-derivative. The R matrix that fits out the least-squares-minimized quadratic is
\begin{equation}
    R = I - M (M^{\mathrm{T}} \Sigma^{-1} M)^{-1} M^{\mathrm{T}}\Sigma^{-1},
\end{equation}
where $I$ is the identity matrix and $M$ is the following design matrix
\begin{equation}
M = 
\begin{bmatrix}
    1 & t_1 & t_1^2 \\
    1 & t_2 & t_2^2 \\
    \vdots & \vdots & \vdots\\
    1 & t_{\mathrm{Ntoas}} & t_{\mathrm{Ntoas}}^2 
\end{bmatrix}.
\end{equation}
\vspace{2mm} 

The columns of $M$ use all TOAs for the pulsar. Finally, $\Sigma$ is the pre-fit noise variance matrix
\begin{equation}
    \Sigma = \langle \mathbf{n}\mathbf{n}^T \rangle.
\end{equation}

The post-fit residuals $\tilde{\boldsymbol{\delta t}}$ and post-fit noise covariance matrix $\tilde{\Sigma}$ then become
\begin{equation}
    \tilde{\boldsymbol{\delta t}} = R \boldsymbol{\delta t} = R (\boldsymbol{\delta t}_{\mathrm{bwm}} + \mathbf{n}) \equiv \tilde{\boldsymbol{\delta t}}_{\mathrm{bwm}} + \tilde{\mathbf{n}},
\end{equation}
with
\begin{equation}
    \tilde{\Sigma} = R \Sigma R^T.
    \label{eqn:postfit_cov}
\end{equation}

We can then write the likelihood for the signal as the probability that the residuals with the signal subtracted out follow the expected Gaussian noise distribution 
\begin{widetext}
\begin{equation}
p(\tilde{\boldsymbol{\delta t}} | \tilde{\boldsymbol{\delta t}}_{\mathrm{bwm}}) = \frac{\exp{\left[ -\frac{1}{2} (\tilde{\boldsymbol{\delta t}} - \tilde{\boldsymbol{\delta t}}_{\mathrm{bwm}})^{\mathrm{T}} \tilde{\Sigma} ^{-1} (\tilde{\boldsymbol{\delta t}} - \tilde{\boldsymbol{\delta t}}_{\mathrm{bwm}})\right]}}{\sqrt{\det{(2 \pi \tilde{\Sigma})}}}.
\label{eqn:marginal_likelihood}
\end{equation}
\end{widetext}

\subsection{General \texorpdfstring{$\pazocal{F}$}{F}-statistic}
The derivation presented in \citet{ellis_optimal_2012} can also be applied to the case of nonlinear gravitational-wave memory. For completeness, we reproduce the steps involved in this derivation.  We begin by defining the inner product between two vectors
\begin{equation}
(\mathbf{x}|\mathbf{y}) \equiv \mathbf{x}^\mathrm{T} \tilde{\Sigma}^{-1} \mathbf{y}.
\label{eqn:inner_prod}
\end{equation}
This allows us to rewrite the log-likelihood (Eq. \ref{eqn:marginal_likelihood}) as
\begin{widetext}
\begin{equation}
\log{p(\tilde{\boldsymbol{\delta t}} | \tilde{\boldsymbol{\delta t}}_{\mathrm{bwm}})} = -\frac{1}{2} (\tilde{\boldsymbol{\delta t}} - \tilde{\boldsymbol{\delta t}}_{\mathrm{bwm}} | \tilde{\boldsymbol{\delta t}} - \tilde{\boldsymbol{\delta t}}_{\mathrm{bwm}}) - \frac{1}{2}\log{\det{(2 \pi \tilde{\Sigma})}},
\label{eqn:compact_likelihood}
\end{equation}
\end{widetext}
where we have suppressed the explicit dependence on the red noise parameters. For  simplicity, in this work we will keep the red noise parameters fixed, and thus the resulting covariance matrix (Eq. \ref{eqn:postfit_cov}) will be constant. Although the previous expressions for the likelihood (Eq \ref{eqn:marginal_likelihood}) are only for a single pulsar, we can easily generalize this to the full PTA case by writing
\begin{equation}
    \tilde{\boldsymbol{\delta t}} = 
    \begin{bmatrix}
        \tilde{\boldsymbol{\delta t}}_1\\
        \tilde{\boldsymbol{\delta t}}_2\\
        \vdots\\
        \tilde{\boldsymbol{\delta t}}_{N_{\mathrm{psr}}}
    \end{bmatrix},
\end{equation}
and
\begin{align}
\tilde{\boldsymbol{\delta t}}_{\mathrm{bwm}} &= 
\begin{bmatrix}
    \tilde{\boldsymbol{\delta t}}_{\mathrm{bwm},1} \\
    \tilde{\boldsymbol{\delta t}}_{\mathrm{bwm},2}\\
    \ldots \\
    \tilde{\boldsymbol{\delta t}}_{{\mathrm{bwm},N_\mathrm{psr}}}
\end{bmatrix}.
\end{align}
In other words, the full residual time series is a vector containing each of the pulsar's timing residual time series. The full $R$ matrix can be written in a block diagonal form
\begin{equation}
    R = 
    \begin{bmatrix}
        R_1 & \mathbf{0} & \cdots & \mathbf{0}\\
        \mathbf{0} & R_2 & \cdots &\mathbf{0}\\
        \vdots & \vdots & \ddots & \mathbf{0} \\
        \mathbf{0} & \mathbf{0} & \cdots & R_{N_{\mathrm{psr}}}
    \end{bmatrix}.
\end{equation}
Finally, for this work, we ignore the small off-diagonal Hellings and Downs spatial correlations that result from the gravitational-wave background and assume the full PTA noise covariance is block-diagonal (see \S~\ref{concl} for further discussion on this assumption)
\begin{equation}
    \tilde{\Sigma} = 
    \begin{bmatrix}
        \tilde{\Sigma}_1 & \mathbf{0} & \cdots & \mathbf{0}\\
        \mathbf{0} & \tilde{\Sigma}_2 & \cdots &\mathbf{0}\\
        \vdots & \vdots & \ddots & \mathbf{0} \\
        \mathbf{0} & \mathbf{0} & \cdots & \tilde{\Sigma}_{N_{\mathrm{psr}}}
    \end{bmatrix}.
\end{equation}
We then write the log-likelihood ratio as
\begin{equation}
\log{\Lambda} = \log{\frac{p(\tilde{\boldsymbol{\delta t}} | \tilde{\boldsymbol{\delta t}}_{\mathrm{bwm}})}{p(\tilde{\boldsymbol{\delta t}} | 0)}} = (\tilde{\boldsymbol{\delta t}}|\tilde{\boldsymbol{\delta t}}_{\mathrm{bwm}}) - \frac{1}{2}(\tilde{\boldsymbol{\delta t}}_{\mathrm{bwm}}|\boldsymbol{\tilde{\delta t}}_{\mathrm{bwm}}),
\label{eqn:loglike_ratio}
\end{equation}
where $p(\tilde{\boldsymbol{\delta t}})$ is the likelihood that these residuals arise from only noise. 

We can write the pre-fit signal template $\boldsymbol{\delta t}_{\mathrm{bwm}}$ as a sum of amplitudes $a_i$ multiplied with template vectors $\mathbf{A}^i$
\begin{equation}
\boldsymbol{\delta t}_{\mathrm{bwm}} = \sum_i{a_i \mathbf{A}^i},
\label{eqn:signal_template_general}
\end{equation}
Just like for the continuous-wave case~\cite{ellis_optimal_2012}, there is more than one way to decompose the signal into the product of an amplitude and template vector. In general, the template vectors $\mathbf{A}^i$ carry the ``ramp"-shaped template, and therefore the time dependence. In this work, Eqs. \ref{eqn:coherent_templates} and \ref{eqn:incoherent_signal_temp} show two different possible choices for template vectors. For now, we will not specify the decomposition used in Eq. \ref{eqn:signal_template_general} but will assume it gives us the full BWM-induced residual time series defined by Eq. \ref{eqn:bwm_residuals} for each pulsar in the full PTA. This will allow us derive general results now that we can apply to the coherent (see \S~\ref{sec:coherent_fstat} ) and incoherent (see \S~\ref{sec:incoherent_fstat}) decompositions later. 

Using Eq.~\ref{eqn:signal_template_general} for the BWM-induced residuals, we can write the log-likelihood ratio (Eq \ref{eqn:loglike_ratio}) as
\begin{equation}
\log\Lambda = a_i (\tilde{\boldsymbol{\delta t}}|\tilde{\mathbf{A}}^i) - \frac{1}{2}a_i a_j(\tilde{\mathbf{A}}^i|\tilde{\mathbf{A}}^j),
\end{equation}
where $\tilde{\mathbf{A}}^i = R \mathbf{A^i}$ are the post-fit time-dependent signal templates. We can then define the matrix elements $\mathbf{N}^i = (\tilde{\boldsymbol{\delta t}}|\tilde{\mathbf{A}}^i)$ and $\mathbf{M}^{ij} = (\tilde{\mathbf{A}}^i|\tilde{\mathbf{A}}^j)$ and maximize this ratio over the amplitude parameters
\begin{equation}
\frac{\partial\log\Lambda}{\partial a_k} = 0 = \mathbf{N}^k - \mathbf{M}^{ik} a_i.
\label{eqn:maximizing_loglike_ratio}
\end{equation}
This gives the maximum likelihood estimator for the amplitude parameters
\begin{equation}
\hat{a}_{i} = \mathbf{M}_{ik} \mathbf{N}^{k},
\end{equation}
with $M_{ik} = (M^{-1})^{ik}$ , and the maximized log-likelihood ratio
\begin{equation}
\log\Lambda = \frac{1}{2} \mathbf{N}^i \mathbf{M}_{ij} \mathbf{N}^j \equiv \pazocal{F}
\end{equation}

The statistic 2$\pazocal{F}$ then follows a $\chi^2$ distribution. As we will show below, depending on our choice of decomposition of the signal, this statistic may be coherent or incoherent. In subsequent sections, we show the choices that give the coherent and incoherent $\pazocal{F}$-statisics, $\pazocal{F}_C$ and $\pazocal{F}_I$, respectively. We note that the dimensions of the $\mathbf{M}$ and $\mathbf{N}$ matrices are determined solely by the number of template vectors. We also show that the coherent statistic $\pazocal{F}_C$ can be computed using two template vectors, with $\mathbf{M}$ a $2\times2$ matrix, and $\mathbf{N}$ a vector of length $2$, whereas the incoherent statistic $\pazocal{F}_I$ only requires one template vector and both $\mathbf{M}$ and $\mathbf{N}$ are scalars.  

\subsection{Coherent \texorpdfstring{$\pazocal{F}$}{F}-statistic} \label{sec:coherent_fstat}
For the coherent statistic, we include the sky-location dependence that comes from the antenna-pattern functions $F^{a}_{+}$ and $F^{a}_{\times}$ in Eq. \ref{eqn:bwm_residuals} as part of the time-dependent template $\mathbf{A}^i$ when we decompose the signal using Eq. \ref{eqn:signal_template_general}.  We write this sky-location dependent time-domain template as 
\begin{align}
\mathbf{C}^{\alpha,m}(\hat{\Omega}, t; t_0) = F_{\alpha,m}(\hat{\Omega}) \Theta(t - t_0) (t-t_0),
\label{eqn:coherent_templates}
\end{align}
where $m=+,\times$ specifies the polarization, and $\alpha$ specifies the pulsar. The corresponding amplitudes are
\begin{align}
    c_+(h_0, \psi) &= h_0 \cos(2\psi), \\
    c_\times(h_0, \psi) &= h_0 \sin(2\psi).
\end{align}
This allows us to rewrite the pre-fit signal template of Eq.~\ref{eqn:signal_template_general} as follows
\begin{align}
    \boldsymbol{\delta t}_{\mathrm{bwm}}(h_0, \psi, \hat{\Omega}, t; t_0) = \sum_{m = +, \times}{c_m}(h_0, \psi) \mathbf{C}^m(\hat{\Omega}, t; t_0),
\end{align}
where
\begin{align}
    \mathbf{C}^m (\hat{\Omega}, t; t_0) &= 
\begin{bmatrix}
    \mathbf{C}^{1,m}(\hat{\Omega}, t; t_0) \\
    \mathbf{C}^{2,m}(\hat{\Omega}, t; t_0) \\
    \ldots \\
    \mathbf{C}^{N_{\mathrm{psr}},m}(\hat{\Omega}, t; t_0)
\end{bmatrix}.
\end{align}
and $\tilde{\mathbf{C}}^{+,\times} = R \mathbf{C}^{+,\times}$ are the post-fit time-dependent signal templates. Following the above derivation, the log-likelihood ratio is maximized for the amplitude parameters
\begin{align}
    \frac{\partial \log \Lambda}{\partial c_k} = 0 &=  (\tilde{\boldsymbol{\delta t}}|\tilde{\mathbf{C}}^i) - c_i(\tilde{\mathbf{C}}^i|\tilde{\mathbf{C}}^k)\\
    &= \mathbf{N}^k - c_i \mathbf{M}^{ik},
\end{align}
where the $\mathbf{M}$ and $\mathbf{N}$ matrices are defined by
\begin{align}
    \left[\mathbf{M}^{ij}\right] &= 
    \begin{bmatrix}
        (\tilde{\mathbf{C}}^+|\tilde{\mathbf{C}}^+) &  (\tilde{\mathbf{C}}^+|\tilde{\mathbf{C}}^\times)\\
         (\tilde{\mathbf{C}}^\times|\tilde{\mathbf{C}}^+) &  (\tilde{\mathbf{C}}^\times|\tilde{\mathbf{C}}^\times)\\
    \end{bmatrix} \\
    \left[\mathbf{N}^i\right] &= 
    \begin{bmatrix}
         (\tilde{\boldsymbol{\delta t}}|\tilde{\mathbf{C}}^+) \\
         (\tilde{\boldsymbol{\delta t}}|\tilde{\mathbf{C}}^\times)
    \end{bmatrix}.
\end{align}
Then, the maximum log-likelihood ratio is
\begin{equation}
    2\pazocal{F}_C = \mathbf{N}^i \mathbf{M}_{ij} \mathbf{N}^j,
\end{equation}
where $\mathbf{M}_{ij}$ is the inverse of $\mathbf{M}^{ij}$. The full expression of $2\pazocal{F}_C$ includes the sum of two correlated random Gaussian variables with unit variance. It is possible to perform a linear transformation to uncorrelate them (\S 3 of Ref. \cite{jaranowski_data_1998} shows this in detail). As such, because $2\pazocal{F}_C$ is computed as the sum of two random Gaussian variables, it follows a $\chi^2$ distribution with two degrees of freedom. In the presence of a signal, it becomes a non-central $\chi^2$-disribution with a non-centrality parameter $\bar{\rho}^2 = (\boldsymbol{\tilde{\delta t}}_{\mathrm{bwm}} | \boldsymbol{\tilde{\delta t}}_{\mathrm{bwm}})$ \cite{jaranowski_data_1998,ellis_optimal_2012}. Note that this non-centrality parameter is also exactly the optimal signal-to-noise ratio (SNR), 
\begin{align}
    \langle 2\pazocal{F}_C \rangle = 2 + \bar{\rho}^2 = 2 + (\boldsymbol{\tilde{\delta t}}_{\mathrm{bwm}} | \boldsymbol{\tilde{\delta t}}_{\mathrm{bwm}}).
    \label{eqn:2fc}
\end{align}
In addition, for the coherent statistic, we can use the maximum likelihood amplitude estimators to easily find the maximum likelihood amplitude parameters for $h_0$ and $\psi$ yielding
\begin{align}
    \hat{\psi} &= \frac{1}{2}\tan^{-1}\left(\frac{\hat{c}_{\times}}{\hat{c}_{+}}\right),
    \label{eqn:mle_psi}
\end{align}
and
\begin{align}
    \hat{h}_0 &= \sqrt{(\hat{c}_{+})^2 + (\hat{c}_{\times})^2}.
    \label{eqn:mle_h}
\end{align}
We note that this calculation can only be done with two or more pulsars. In the case that the PTA consists of only a single pulsar, the matrix $\mathbf{M}_{ij}$ is non-invertible. This is expected, since a single pulsar cannot simultaneously constrain the GW strain $h_0$ and polarization $\psi$.
\subsection{Incoherent \texorpdfstring{$\pazocal{F}$}{F}-statistic}\label{sec:incoherent_fstat}
The other possibility is to decompose the signal using the time-domain signal template $\mathbf{D}$
\begin{equation}
    \mathbf{D}= 
    \begin{bmatrix}
    \mathbf{D}^1(t; t_0) \\
    \mathbf{D}^2(t; t_0) \\
    \vdots \\
    \mathbf{D}^{N_{\mathrm{psr}}}(t; t_0)
    \end{bmatrix},
\end{equation}
where each element of the column vector is the time-dependent template for a single pulsar
\begin{equation}
\mathbf{D}^\alpha(t, t_0) = \Theta(t-t_0) (t-t_0),   
\end{equation}
where again $\Theta(t-t_0)$ is the Heaviside function. The corresponding amplitude for a pulsar indexed by $\alpha$ is
\begin{equation}
d_\alpha(h_0, \psi, \hat{\Omega}) = h_0 (\cos(2\psi) F_\alpha^{+}(\hat{\Omega}) + \sin(2\psi)F_\alpha^\times(\hat{\Omega})).
\label{eqn:fi_amp}
\end{equation}
The post-fit signal for the entire PTA can then be expressed as
\begin{equation}
\tilde{\boldsymbol{\delta t}}_{\mathrm{bwm}}(h_0, \psi, \hat{\Omega}, t; t_0) = 
\begin{bmatrix}
    d_1(h_0, \psi, \hat{\Omega}) \tilde{\mathbf{D}}^1(t;t_0) \\
    d_2(h_0, \psi, \hat{\Omega}) \tilde{\mathbf{D}}^2(t;t_0) \\
    \vdots \\
    d_{N_{\mathrm{psr}}}(h_0, \psi, \hat{\Omega}) \tilde{\mathbf{D}}^{N_{\mathrm{psr}}}(t;t_0)
\end{bmatrix}.
\label{eqn:incoherent_signal_temp}
\end{equation}
where again we use the post-fit time-dependent template $\tilde{\mathbf{D}} = R\mathbf{D}$. 
The log-likelihood ratio for a memory signal to the null-signal model is
\begin{align}
    \log{\Lambda} &= (\tilde{\boldsymbol{\delta t}}|\tilde{\boldsymbol{\delta t}}_{\mathrm{bwm}}) - \frac{1}{2}(\tilde{\boldsymbol{\delta t}}_{\mathrm{bwm}}|\tilde{\boldsymbol{\delta t}}_{\mathrm{bwm}}) \\
    &= \sum_{\alpha}^{N_{\mathrm{psr}}}{\left[(\tilde{\boldsymbol{\delta t}}^{\alpha}|\tilde{\boldsymbol{\delta t}}^{\alpha}_{\mathrm{bwm}}) - \frac{1}{2}(\tilde{\boldsymbol{\delta t}}^{\alpha}_{\mathrm{bwm}}|\tilde{\boldsymbol{\delta t}}^{\alpha}_{\mathrm{bwm}})\right]} \\
    &= \sum_{\alpha}^{N_{\mathrm{psr}}}{\left[d_{\alpha}(\tilde{\boldsymbol{\delta t}}^{\alpha}|\tilde{\mathbf{D}}^{\alpha}) - \frac{1}{2} d_{\alpha}^2(\tilde{\mathbf{D}}^{\alpha}|\tilde{\mathbf{D}}^{\alpha})\right]} .\label{eqn:incoh_loglike_sum}
\end{align}
Maximizing over the $\beta$-th amplitude gives
\begin{equation}
    \frac{\partial\log\Lambda}{\partial d_\beta} = 0 =  (\tilde{\boldsymbol{\delta t}}^{\beta}|\tilde{\mathbf{D}}^{\beta}) - d_{\beta}(\tilde{\mathbf{D}}^{\beta}|\tilde{\mathbf{D}}^{\beta})
\end{equation}
Solving yields the maximal likelihood estimator for the $\beta$-th amplitude parameter
\begin{equation}
    \hat{d}_{\beta} = \frac{(\tilde{\boldsymbol{\delta t}}^{\beta}|\tilde{\mathbf{D}}^{\beta})}{(\tilde{\mathbf{D}}^{\beta}|\tilde{\mathbf{D}}^{\beta})} = \mathbf{N}_\beta \mathbf{M}^{-1}_\beta
\end{equation}
with
\begin{align}
    \mathbf{M}_{\beta} = (\tilde{\mathbf{D}}^{\beta}|\tilde{\mathbf{D}}^{\beta})\\
    \mathbf{N}_{\beta} = (\tilde{\boldsymbol{\delta t}}|\tilde{\mathbf{D}}^{\beta}),
\end{align}
where $\mathbf{M}_\beta$ and $\mathbf{N}_\beta$ are both scalars since there is only one amplitude and one template in this factorization of the signal.

The maximal log-likelihood (which we define as $\pazocal{F}_I$) is then Eq. \ref{eqn:incoh_loglike_sum} after substituting the maximum likelihood estimates for the amplitudes
\begin{align}
    \pazocal{F}_I &= \sum_{\alpha}^{N_{\mathrm{psr}}}{\left[\hat{d}_{\alpha}(\tilde{\boldsymbol{\delta t}}^{\alpha}|\tilde{\mathbf{D}}^{\alpha}) - \frac{1}{2} \hat{d}_{\alpha}^2(\tilde{\mathbf{D}}^{\alpha}|\tilde{\mathbf{D}}^{\alpha})\right]}\\
    &= \sum_{\alpha}^{N_{\mathrm{psr}}}{\left[\frac{1}{2}\frac{\mathbf{N}_\alpha^2}{\mathbf{M}_\alpha}\right]}.
\end{align}
Finally,
\begin{equation}
    2\pazocal{F}_I = \sum_{\alpha}^{N_{\mathrm{psr}}}{\frac{\mathbf{N}_\alpha^2}{\mathbf{M}_\alpha}}
\end{equation}

Because $2\pazocal{F}_I$ is a sum of $N_{\mathrm{psr}}$ independent Gaussian random variables with unit variance, it follows a $\chi^2$ distribution with $N_{\mathrm{psr}}$ degrees of freedom. However, the non-centrality parameter $\bar{\rho}^2 = (\boldsymbol{\tilde{\delta t}}_{\mathrm{bwm}} | \boldsymbol{\tilde{\delta t}}_{\mathrm{bwm}})$ remains the same as the non-centrality of the $2\pazocal{F}_C$ statistic
\begin{equation}
\langle 2 \pazocal{F}_I \rangle = N_{\mathrm{psr}} + \bar{\rho}^2 = N_{\mathrm{psr}} +(\boldsymbol{\tilde{\delta t}}_{\mathrm{bwm}} | \boldsymbol{\tilde{\delta t}}_{\mathrm{bwm}}).
\label{eqn:2fi}
\end{equation}

As in the case for the $\pazocal{F}_C$ statistic it is possible to solve for the signal parameters using the maximum likelihood estimators for the amplitude. Since there are four independent parameters in the amplitudes $\hat{d}_{\beta}$ (see Eq. \ref{eqn:fi_amp}), it is possible to numerically solve for the maximum likelihood estimators by using a minimum of four pulsars.

\section{Methodology and Results}
\subsection{Simulated Data Sets}
To test our statistics we perform two different sets of simulations. First, we simulate PTAs with 40 pulsars placed in random, uniformly distributed sky locations, with time baselines of 10 years. Each pulsar timing residual dataset is composed of Gaussian white noise, power-law red noise, and a BWM signal. The signal source is placed at the average of the random right-ascensions and declinations of all the pulsars in the PTA. The injected Gaussian white noise has an amplitude of $\sigma_{\mathrm{WN}} = 100 \mathrm{ns}$. The red noise (following Eq. \ref{eqn:powerlaw}) has an amplitude of $A=3.0\times10^{-15}$ and a spectral index $\gamma = \frac{13}{3}$. Finally, the memory signal is constructed with a log-strain $\log{h_0} = -13.5$, burst epoch $t_0 = \frac{2}{5} T_{\mathrm{PTA}}$, and a polarization angle $\psi = 0$. These injection parameters are summarized in Table \ref{tab:fstat_injection_params}. While red noise from a stochastic gravitational wave background has quadrupolar spatial correlations, we exclude these in this work for simplicity. The red noise injected in these data have the same power spectrum in each pulsar and are spatially uncorrelated. This common uncorrelated red noise is often called CURN. We then create $5000$ realizations for each combination of noises and signal, and compute both $\pazocal{F}_I$ and $\pazocal{F}_C$. The results are shown in Figure \ref{fig:fstat_recovery}.

\begin{table*}
    \begin{tabular}{|l|c|c|}\hline
        \textbf{Parameter Name} & \textbf{Variable} & \textbf{Injected Value} \\ \hline \hline
        Gaussian White Noise Amplitude & $\sigma_{\mathrm{WN}}$ & $100$ ns \\ \hline
        Red Noise Amplitude & $A_{\mathrm{RN}}$ & $3\times10^{-15}$ \\ \hline
        Red Noise Spectral Index & $\gamma$ & ${13}/{3}$ \\ \hline
        Memory Strain Amplitude & $h_0$ & $3.16\times10^{-14}$ \\ \hline
        Memory Epoch & $t_0$ & $4$ yr \\ \hline
        Memory Polarization & $\psi$ & $0$ rad \\ \hline
        
    \end{tabular}
    \caption{This table summarizes the values of the injected noise and signal parameters used to create $5000$ realizations of simulated data sets. These realizations were used to compute the $\pazocal{F}$-statistics shown in Figure \ref{fig:fstat_recovery}.} \label{tab:fstat_injection_params}
\end{table*}
We also show the parameter estimation capabilities of the $\pazocal{F}_C$ statistic using Eqs. \ref{eqn:mle_psi} and \ref{eqn:mle_h}. 

To do this, we simulate data sets with the same white noise ($\sigma_{\mathrm{wn}} = 100 \mathrm{ns}$) and red noise ($A_{\mathrm{RN}} = 3.0\times10^{-15}, \gamma = {13}/{3}$), and inject signals at three different strengths $A_{\mathrm{bwm}} = \{5\times10^{-15}, 8.5\times10^{-15}, 1.5\times 10^{-14}\}$ at a polarization angle of $\psi = {\pi}/{8} \sim 0.39$ rad.

Table \ref{tab:recovery_table} shows the injected signal parameters as well as the results of these parameter estimations. We also present 2-dimensional histograms of the recovered signal parameters in Figures \ref{fig:maxlike_param_recovery_weak}-\ref{fig:maxlike_param_recovery_strong}.

\subsection{Results and Discussion}
\begin{figure*}
    \centering
    \includegraphics[width=0.88\textwidth]{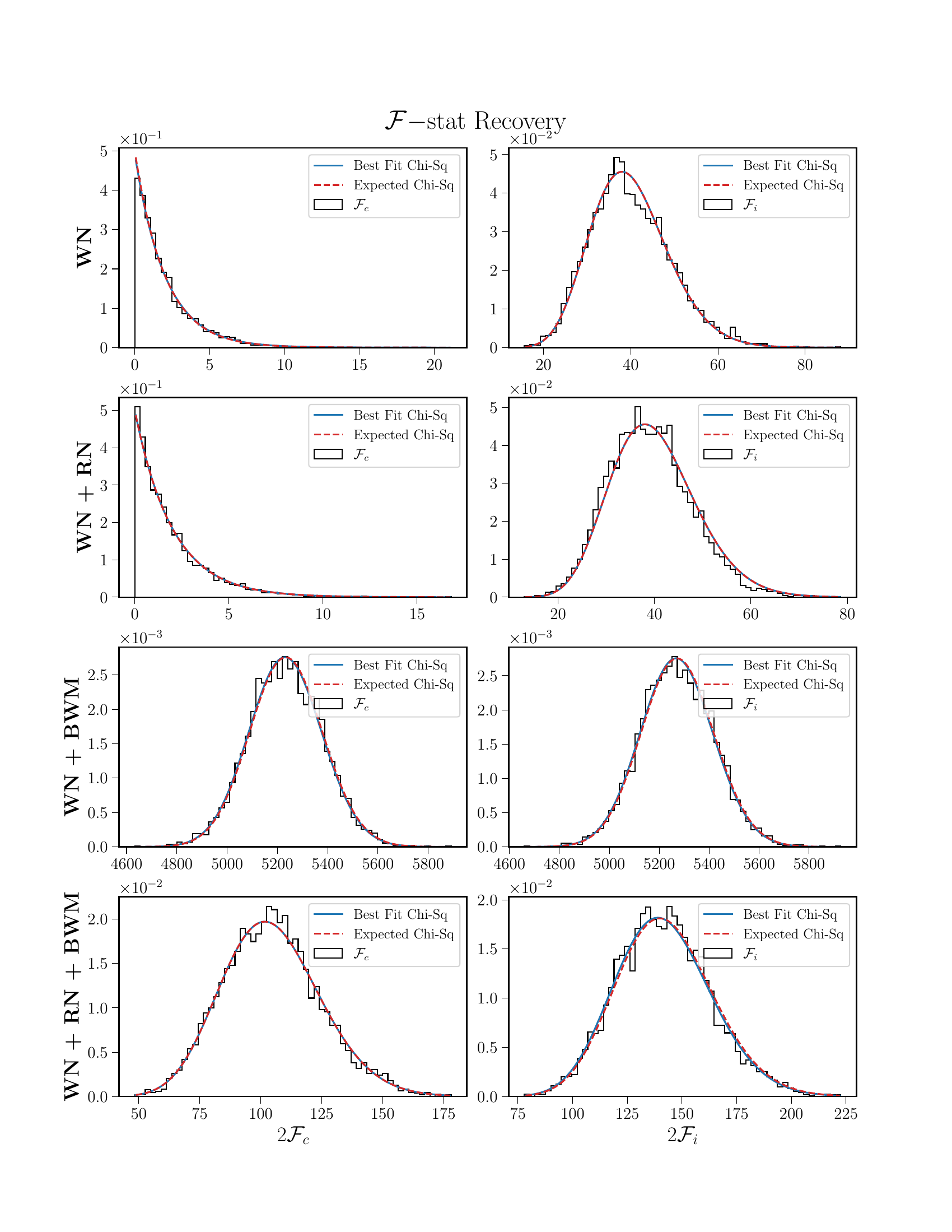}
    \caption{This figure shows the best-fit $\chi^2$-distributions to the $\pazocal{F}_C$ and $\pazocal{F}_I$ statistics in multiple scenarios containing different combinations of gaussian white noise, a common spatially uncorrelated red noise (CURN), and a nonlinear memory signal. 
    We see that in these cases, the theoretically predicted distributions and the theoretical non-centrality parameter match almost exactly with the distributions of the $\pazocal{F}$-statistics from $5000$ simulated realizations. Notably, the SNR decreases significantly when comparing the statistics in the presence of red noise.}
    \label{fig:fstat_recovery}
\end{figure*}

\begin{table*}
    \begin{tabular}{|l|c|c|c|c|c|} \hline
         \textbf{Signal Regime} & \textbf{SNR} & \multicolumn{2}{|c|}{\textbf{Injected Parameters}} & \multicolumn{2}{|c|}{\textbf{Recovered Parameters}} \\ \hline \hline
         & & $\log_{10}{h_0}$ & $\psi$ & $\log_{10}{\hat{h}_0}$ & $\hat{\psi}$ \\ \hline
        Weak & $1.74$ & $-14.30$ & $0.39$ & $-14.19^{+0.20}_{-0.28}$ & $0.29^{+0.31}_{-0.67}$ \\ \hline
        Intermediate & $5.03$ & $-14.07$ & $0.39$ & $-14.03^{+0.16}_{-0.21}$ & $0.38^{+0.20}_{-0.33}$ \\ \hline
        Strong & $15.67$ & $-13.82$ & $0.39$ & $-13.81^{+0.11}_{-0.13}$ & $0.39^{+0.12}_{-0.15}$ \\ \hline
    \end{tabular}
     \caption{This table shows the signal-to-noise ratio, the injected signal parameters, and maximum-likelihood estimators for the signal parameters recovered using $\pazocal{F}_C$ in multiple signal regimes over $5000$ realizations of simulations. These regimes are determined based on the SNR of the signal compared to the injected noise, which was kept constant ($\sigma_{\mathrm{WN}} = 100$ ns, $A_{\mathrm{RN}} = 3\times10^{-15}$, $\gamma = 13/3$). The maximum-likelihood estimators for the injected memory strain-amplitude is very imprecise for low-SNR signals, but becomes much more accurate at higher SNR. This is because GW memory signal has a similar power spectrum to a red noise with $\gamma = 13/3$. However, the background is loud compared to the GW memory signal expected from a modest SMBHB merger. The sensitivity of the PTA to memory thus has a lower limit determined by the amplitude of the red noise.} \label{tab:recovery_table}
\end{table*}

\begin{figure*}
    \centering
    \subfloat[]{
        \includegraphics[width=0.42\textwidth]{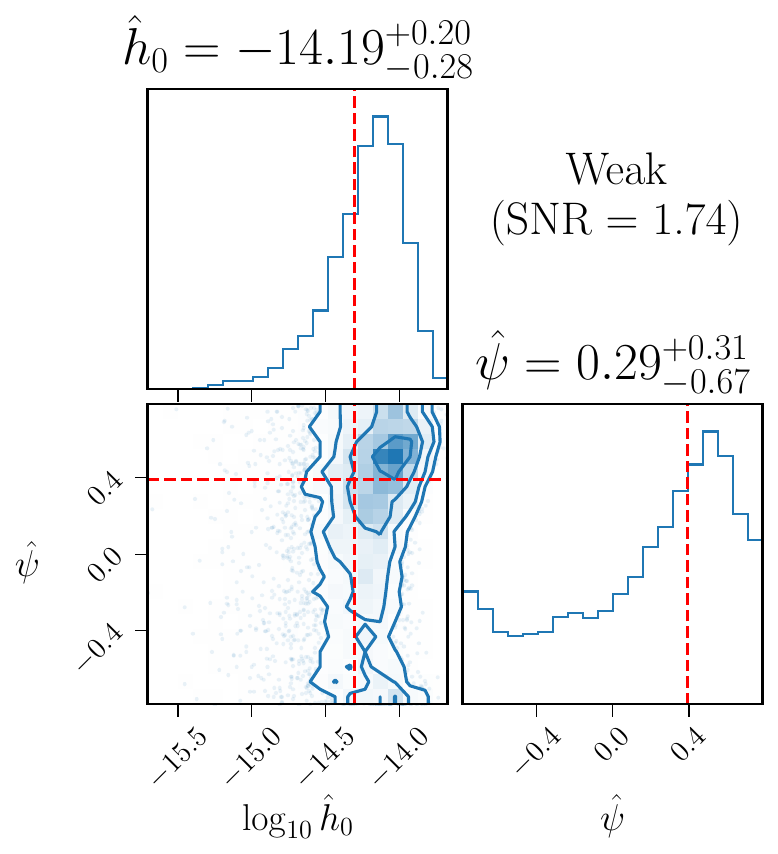}
        \label{fig:maxlike_param_recovery_weak}
    }\quad
    \subfloat[]{
        \includegraphics[width=0.42\textwidth]{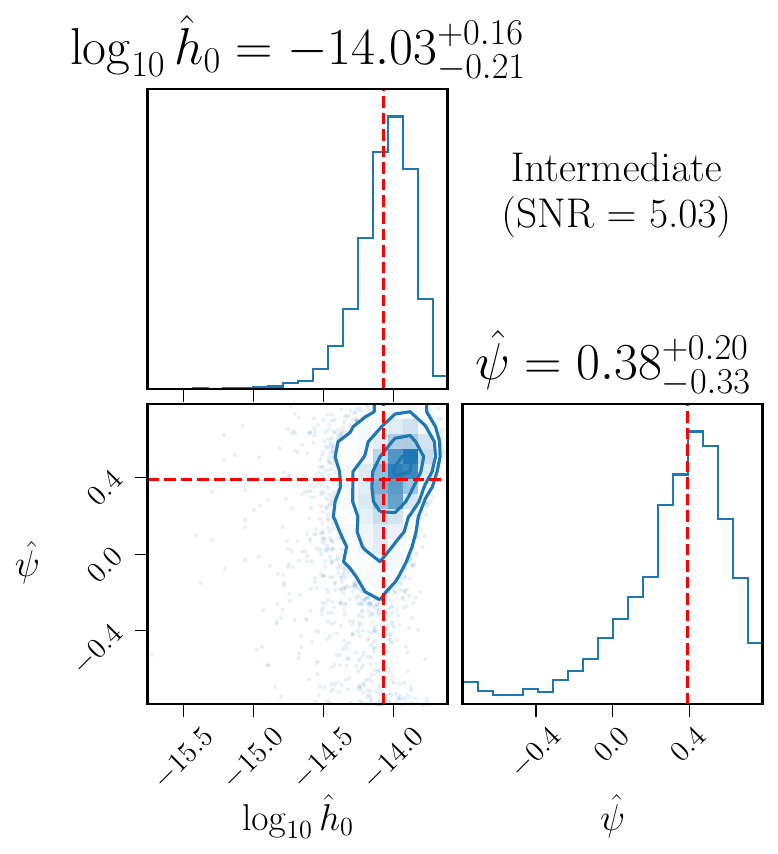}
        \label{fig:maxlike_param_recovery_intermediate}  
    }\quad
    \subfloat[]{
        \includegraphics[width=0.42\linewidth]{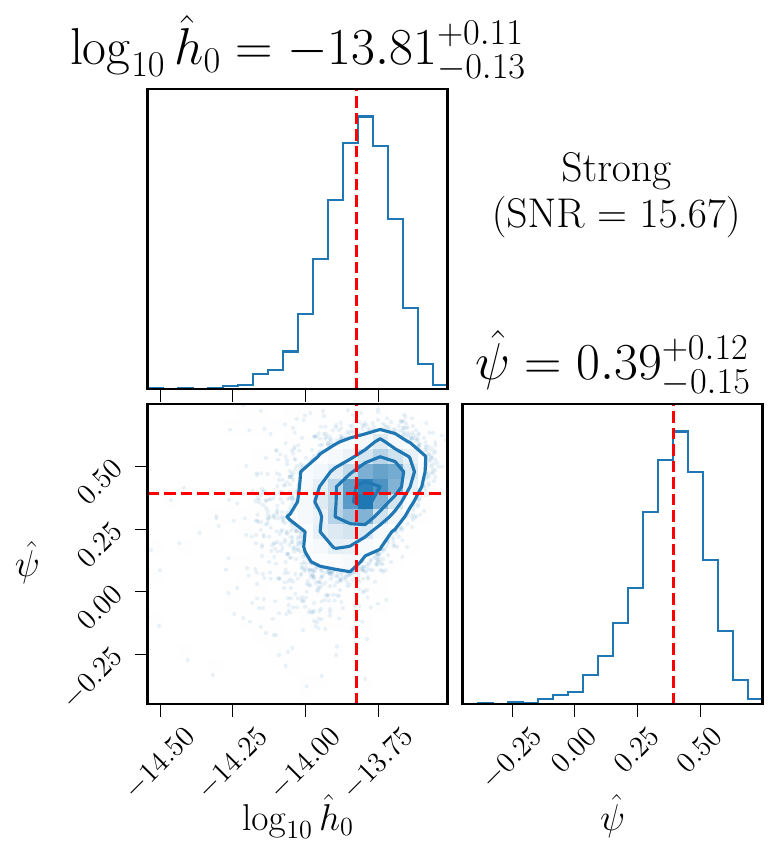}
        \label{fig:maxlike_param_recovery_strong}
    }
    \label{fig:maxlike_param_recoveries}
    \caption{This figure shows the results of maximum-likelihood parameter recovery using Eqs. \ref{eqn:mle_psi} and \ref{eqn:mle_h} in multiple signal regimes. The summary of the injection and recoveries may be found in Table \ref{tab:recovery_table}. As expected, the maximum likelihood estimators for the strain amplitude $h_0$ and $\psi$ do very poorly at small SNR (Figure \ref{fig:maxlike_param_recovery_weak}, SNR $= 1.74$). However, with a moderately large signal (Figure \ref{fig:maxlike_param_recovery_intermediate}, SNR $= 5.03$), the strain amplitude estimator becomes more accurate, while the polarization is not well measured in some realizations. Only when the signal is very loud relative to the noise (Figure \ref{fig:maxlike_param_recovery_strong}, SNR $= 15.67$) does the maximum likelihood estimator for the $\psi$ become accurate. The weak sensitivity to polarization angle is largely due to the total size (40 pulsars) and distribution (uniform) of the pulsars in this PTA. A PTA with more pulsars distributed more densely near the sky location of the signal source would better constrain the polarization of the signal.}
\end{figure*}

We compute both the $\pazocal{F}_C$ and $\pazocal{F}_I$ statistics for each combination of injected noise and signals over 5000 realizations of the noise processes. To compare the results with analytic expectations we calculate the non-centrality parameters $\bar{\rho}^2$ as defined in Eqs. \ref{eqn:2fc} and \ref{eqn:2fi}, and we use them to show the consistency between the best-fit $\chi^2$-distributions and the theoretically predicted $\chi^2$-distributions in Figure \ref{fig:fstat_recovery}. 

In all cases where there is no signal present, the top four plots in Fig.~\ref{fig:fstat_recovery}, the post-fit covariance matrix properly accounts for both red and white noise. In these cases, the best-fit $\chi^2$-distribution is nearly identical to the expected null-signal $\chi^2$-distribution. When a signal is present in our data, bottom four plots in Fig.~\ref{fig:fstat_recovery}, the best-fit non-centrality parameter is in excellent agreement with analytic predictions.  

Notably, the addition of red noise into a data set with a signal significantly decreases the non-centrality parameter, which is just the optimal SNR. This is consistent with our expectations, since our BWM signal (which is just a ramp in the time domain) is partially covariant with powerlaw red noise processes. We find that the non-centrality parameter is reduced from approximately $\sim5000$ in the case of signal and white noise to $\sim150$ in the case of signal, white noise, and red noise. This reduction in the non-centrality parameter means that --- even using the exact values of the injection parameters to compute the $\pazocal{F}-$statistics --- our ability to the detect a BWM signal is hampered significantly by the presence of red noise. This corroborates the results from refs. \cite{cordes_detecting_2012, madison_assessing_2014} regarding the effects of red noise on the detectability of GW memory in PTAs.

We can also see this degeneracy between red noise and nonlinear GW memory in Figure \ref{fig:maxlike_param_recovery_weak}. In these parameter estimation runs, we attempt to recover the injected signal parameters using the maximum-likelihood estimators in Eqs. \ref{eqn:mle_psi} and \ref{eqn:mle_h}. The injected and recovered strain-amplitudes and polarizations are summarized in Table \ref{tab:recovery_table}. The 2-dimensional histograms of the estimated maximum-likelihood parameters are also shown in Figure \ref{fig:maxlike_param_recovery_weak} through Figure \ref{fig:maxlike_param_recovery_strong}. 

As seen in the figures, as the injected signal amplitude becomes weaker, parameter recovery becomes less accurate. In the weak-signal case (Figure \ref{fig:maxlike_param_recovery_weak}), the amplitude of the memory is small compared to the red noise amplitude. We see that the median recovered maximum likelihood amplitude ($\log_{10}{\hat{h}_0} = -14.19$) is larger than the injected value ($\log_{10}{h_{0, \mathrm{inj}}} = -15$), and the polarization angle is not well recovered in many realizations. This is because the memory signal is completely hidden by the red noise. Aas a result, the coherent statistic, which makes use of the correlations between pulsars in the antenna response to the signal, cannot correctly recover the injected signal parameters. Thus, red noise sets a high noise ``floor" for GW memory detection.

In the intermediate case (Figure \ref{fig:maxlike_param_recovery_intermediate}), we again see that the maximum likelihood estimator for the amplitude of the memory signal  ($\log_{10}{\hat{h}_{0}} = -14.03$) estimates the injected strain-amplitude accurately ($\log_{10}{{h}_{0}} = -14.07$) for many realizations. The estimator for the polarization is improved, but still inaccurate in some realizations.  

In the case of very strong signals (Figure \ref{fig:maxlike_param_recovery_strong}), we see that the median maximum-likelihood estimators over all realizations are more accurate. To obtain an order-of-magnitude estimate of the size of the merging objects and provide a sense of scale, we can use Eq. 1 in \cite{madison_assessing_2014}
\begin{equation}
h_{0} = \frac{1-\sqrt{8}/3}{24} \frac{G\mu}{c^2 r}\sin^2{\iota}(17 + \cos^2{\iota}),
\label{eqn:madison_strain_amplitude}
\end{equation}
where $G$ is the gravitational constant, $\mu$ is the reduced mass of the SMBHB merger system, $c$ is the speed of light, $r$ is the comoving distance to the SMBHB, and $\iota$ is the inclination angle of the binary. At a fiducial comoving distance of $r = 1$ Gpc, the strain amplitudes in the intermediate- and strong- regimes, $h_{\mathrm{intermediate}} = 8.5\times10^{-15}$, and $h_{\mathrm{strong}} = 1.5\times10^{-14}$,  correspond to the merger of two equal-mass SMBHs of  $2.7\times10^9$ \(M_\odot\) and $4.8\times 10^9$ \(M_\odot\), respectively. 


\section{Conclusion}
\label{concl}
We have presented the derivation of two statistics analogous to the continuous GW $\pazocal{F}$-statistic for use in the detection of nonlinear GW memory in PTA data. These statistics are computed by analytically maximizing the likelihood over the amplitude parameters of the nonlinear GW memory signal model. This may be done in two different ways: coherently, in which case we fix a source sky location and use the resulting antenna response of the PTA in the construction of the time-domain templates; or incoherently, in which case we do not use the antenna response of the PTA as part of the template and maximize over sky-locations along with the strain and polarization of the GW.  The coherent statistic $\pazocal{F}_C$ is the analog of the continuous wave $\pazocal{F}_e$, and the incoherent statistic $\pazocal{F}_I$ is the analog of $\pazocal{F}_p$ \cite{ellis_optimal_2012}.  

We have shown that in simple data sets including both Gaussian white and red noise, both statistics follow the expected non-central $\chi^2$-distributions with the non-centrality parameters given by the optimal SNR $\bar{\rho}^2 = (\boldsymbol{\tilde{\delta t}}_{\mathrm{bwm}} | \boldsymbol{\tilde{\delta t}}_{\mathrm{bwm}})$. We have also demonstrated the parameter-estimation capabilities of the coherent statistic. As expected, at low SNR ($\mathrm{SNR} \lesssim 5$), the maximum likelihood estimators for both the strain amplitude and polarization of the nonlinear memory signal are inaccurate. In the intermediate SNR regime, only the amplitude estimator becomes reliably accurate. The polarization estimator only becomes accurate at very high SNR. This can be mitigated by increasing the number of pulsars in the PTA, since increased sky coverage allows the PTA to rule out more source orientations.


We are currently working to incorporate the the $\pazocal{F}_C$ and $\pazocal{F}_I$  statistics into the standard pulsar timing analysis software extension \texttt{enterprise\textunderscore extensions} package \footnote{\url{https://github.com/nanograv/enterprise_extensions}}\citep{enterprise}. This will allow the statistics to be used in future searches for nonlinear GW memory in PTA data sets. 
Compared with a full Bayesian search, the $\pazocal{F}_C$ and $\pazocal{F}_I$ statistics are efficient because we maximize, rather than marginalize, over signal parameters. However, we anticipate that as with other frequentist methods it will be necessary to develop noise-marginalized versions of these statistics to properly account for the broad posterior distributions of the pulsars' intrinsic noise parameters and avoid potential biases (such as those found in the optimal statistic for the GWB \cite{vigeland_noise-marginalized_2018}). Despite the need for noise marginalization we still expect this procedure to be significantly more efficient than a full Bayesian search, similar to the case of the noise-marginalized optimal statistic for the stochastic GW background \cite{vigeland_noise-marginalized_2018}.

In future studies we will also use these statistics on more realistic datasets to better understand the impacts of different intrinsic red and white noises for each pulsar, different time baselines for different pulsars, searching over source sky locations, and correlations of the common red noise process that will more accurately model the stochastic gravitational-wave background. 



\section{Acknowledgments}
We would like to thank Jeremy Baier and members of the NANOGrav Detection Working group for useful comments on the manuscript. JPS, XS, and DRM are members of the NANOGrav Collaboration. The NANOGrav collaboration receives support from National Science Foundation (NSF) Physics Frontiers Center award No. 2020265. This work is also partly supported by the George and Hannah Bolinger Memorial Fund in the College of Science at Oregon State University. 

%




\pagebreak
\bibliography{bibliography}
\bibliographystyle{aasjournal}



\end{document}